# Assessment of Mudrock Brittleness with Micro Scratch Testing

Luis Alberto Hernandez-Uribe, Michael D. Aman, D. Nicolas Espinoza*

*Corresponding Author: espinoza@autin.utexas.edu

Department of Petroleum and Geosystems Engineering

The University of Texas at Austin, United States

200 E. Dean Keeton St., Austin, TX 78712, USA

**Abstract**

Mechanical properties are essential for understanding natural and induced deformational behavior of geological formations. Brittleness characterizes energy dissipation rate and strain localization at failure. Brittleness has been investigated in hydrocarbon-bearing mudrocks in order to quantify the impact of hydraulic fracturing on the creation of complex fracture networks and surface area for reservoir drainage. Typical well logging correlations associate brittleness with carbonate content or dynamic elastic properties. However, an index of rock brittleness should involve actual rock failure and have a consistent method to quantify it. Here we present a systematic method to quantify mudrock brittleness based on micro-mechanical measurements from the scratch test. Brittleness is formulated as the ratio of energy associated with brittle failure to the total energy required to perform a scratch. Soda lime glass and polycarbonate are used for comparison to identify failure in brittle and ductile mode and validate the developed method. Scratch testing results on mudrocks indicate that it is possible to use the recorded transverse force to estimate brittleness. Results show that tested samples rank as follows in increasing degree of brittleness: Woodford, Eagle Ford, Marcellus, Mancos, and Vaca Muerta. Eagle Ford samples show mixed ductile/brittle failure characteristics. There appears to be no definite correlation between micro-scratch brittleness and quartz or total carbonate content. Dolomite content shows a stronger correlation with brittleness than any other major mineral group. The scratch brittleness index correlates positively with increasing Young's modulus and decreasing Poisson's ratio, but shows deviations in rocks with distinct porosity and with stress-sensitive brittle/ductile behavior (Eagle Ford). The results of our study demonstrate that the micro-scratch test method can be used to investigate mudrock brittleness. The method is particularly useful for reservoir characterization methods that take advantage of drill cuttings or whenever large samples for triaxial testing or fracture mechanics testing cannot be recovered.

# 1. INTRODUCTION

Natural gas production in the United States is expected to increase at an annual average rate of 1.8% over the next 25 years, mostly as a result of continued development of shale drilling and completion (EIA, 2016). The current prominence of shale gas production has in large part resulted from the use of hydraulic fracturing, a well stimulation technique which uses pressure and proppants to open fractures (Hubert and Willis, 1972, Cosgrove, 2005; Harrison, 2005; Kumar et al., 2012). (). As conventional reservoirs become increasingly sparse it may be expected that techniques including hydraulic fracturing become more common in unconventional reservoir production.

A large fraction of unconventional hydrocarbon production comes from organic mudrock, a sedimentary rock rich in fine particles and organic matter (Bell, 2005; EIA, 2016,). Brittleness plays a key role in hydraulic fracturing of mudrocks (Bai, 2016). Brittle rocks tend to break dissipating energy in a short period of time and creating large surface area upon failure, while ductile rocks tend to absorb large amounts of energy, by means of plastic work, and distribute strains when loaded (Bell 2005; Jaeger et al., 2007). Understanding and quantifying brittleness in a geological formation is essential in processes like hydraulic fracturing which seeks to increase reservoir surface area connected to the wellbore (Jin et al., 2014; Sharma and Manchanda, 2015; Bai, 2016).

Brittleness is a fundamental material property also observed in ceramics, metals, and polymers and has several definitions depending on the field of study (Cottrell, 1958; Lawn, 1993; Brostow et al., 2006). Rock mechanics uses several definitions of brittleness based on elastic and plastic strains, compression and tensile strength, and post-peak behavior (Hucka and Das, 1974; Holt et al., 2011; Yang et al., 2013; Meng et al., 2015; Bai, 2016). Wellbore geophysics often makes use of small-strain dynamic elastic parameters and mineralogical content -mostly quartz, carbonates, feldspars, and mica- to develop correlations to predict rock brittleness (Jarvie et al., 2007; Wang and Gale, 2009; Holt et al., 2011; Jin et al., 2014; Herwanger and Mildren, 2015).

This paper presents a new method to investigate the brittleness of mudrocks using the micro-scratch test method. The objective of this work is to present a method that can determine rock matrix brittleness based on actual failure phenomena rather than mineralogical correlations or small-strain elastic parameters. We show experimental results and analysis on five different organic mudrocks. Further analysis compares calculated brittleness with expectations from mineralogical content and elastic parameters.

## 2. BACKGROUND: MICRO-SCRATCH TEST METHOD AND BRITTLENESS

Micro-scratch tests use a probe-tipped device to create a scratch. The test consists of measuring the transverse force and position of the probe tip as a scratch is made on a flat surface. Surface roughness should be five times less than the average scratch penetration depth (Miller et al., 2008). A constant normal force is applied to the probe tip as the sample is scratched at a constant speed (another variant applies fixed scratch depth). Based on the measurement of the applied forces and scratch size, various mechanical properties may be calculated, such as material hardness, UCS and fracture toughness (ASTM G171; Akono and Ulm, 2011; Richard et al., 2012; Sun et al., 2016). Scratch testing can result in ductile and brittle failure mechanisms. Figure 1 shows scratches made on soda lime glass and on polycarbonate. Visual examination of the scratch surface permits identifying these contrasting failure modes: brittle (in the case of soda lime glass) and ductile (in the case of polycarbonate).

Brittle failure is associated with microscopic fracture events where cracks propagate from the probe tip and pieces of material disengage in chunks leaving a set of aligned sharp wedges – see Fig. 1a (Huang and Detournay, 2008; Akono et al., 2011; Akono et al., 2012). Hence, instances of increasing transverse force may correspond with accumulation of energy stored in the material domain near the probe prior to brittle failure or "chipping" (Akono and Ulm, 2011). Acoustic emission monitoring shows that stress induced damage events often correspond to sudden drops in horizontal force while making a scratch in brittle materials (Akono and Ulm, 2011; Akono and Kabir, 2015). Brittle chipping may not develop below a threshold scratch depth in a brittle material (Richard et al., 1998; Schei et al., 2000).

On the other hand, ductile failure results in a smooth scratch surface as a result of plastic "flow" around the tip or de-cohesion of grains in the case of poorly cemented rocks (Richard et al., 1998; Schei et al., 2000; Richard et al., 2012). In ductile deformation, de-cohesion of matrix material may occur and grains and powder may accumulate in front of the scratch probe and on the sides of residual scratches (Richard et al., 1998; Richard et al., 2012). Plastic flow does not exhibit instances of rapid strain energy release. However, instances of increasing transverse force may correspond with ductile failure while the scratch depth deepens.

Brittle and ductile failure have been observed to possess distinct transverse force signals. Whereas brittle failure has a pronounced "saw-tooth" transverse force response with large fluctuations, ductile failure has a profile with smaller fluctuations (Richard et al., 2012). Figure 1 confirms distinct transverse force signatures involved in scratching ductile and brittle materials. Photos of scratch residuals demonstrate that soda lime glass scratch is abundant in sharp fracture planes whereas the polycarbonate scratch surface is entirely smooth.

## 3. EXPERIMENTAL PROGRAM

### 3.1 Scratch Apparatus

The micro-scratch assembly is composed of a sample fixture, stylus with scratch tip, a stepper motor, and displacement and load transducers ([Figure 2a](#)). A vise is used to secure a sample while a normal and transverse force are applied. The vise is mounted to an XY stage (Newport Corporation 9064) so it is free to move in the horizontal direction when a transverse force is applied. Scratches were made with a spherical-conical diamond probed-tip with a tip radius of 200 µm and cone angle of 120° (Gilmore Diamond Tools, Inc.) stationed in a brass collar above the vise. The conical-spherical tip is free to move in the vertical direction. [Figure 2b](#) shows the conical-spherical tip in contact with a mudrock sample. While the scratch tip and sample are in contact, a stepper motor system is used to move the stage and scratch the sample. The transverse force applied to the sample is measured by a Pasco PS-2200 100 N load cell, and the position of the sample is measured by an Omega LDI-119 linear variable differential transformer (LVDT). An Arduino UNO REV3 board controls the NEMA 17 stepper motor, and a Keysight 34972A data acquisition unit records transducer signals with time. Scratches are photographed with a Nikon SMZ800N stereomicroscope.

### 3.2 Samples Investigated and Preparation

This study investigates organic mudrocks as well as polycarbonate and soda lime glass as controls. The polycarbonate tested is Lexan 9034, manufactured by Regal Plastics. Organic mudrocks investigated include samples from Eagle Ford, Mancos, Marcellus, Vaca Muerta, and Woodford formations. Rock samples came from pieces of field cores and outcrops ([Table 1](#)). Only Eagle Ford came as a large sample from which multiple samples could have been cut. Samples were cut into rectangular prisms of 1 inch by 1 inch with a height of 0.25 inches with a diamond wafering saw. Sample surfaces were then sanded with successively smaller grit sandpaper and then polished with Allied Tech diamond lapping film on a lapping wheel to 9 µm grade film.

We measured dynamic elastic moduli with ultrasonic P-wave and S-wave transducers developed by Terratek at a frequency around 1 MHz. [Table 1](#) summarizes the mass density, dynamic Young's modulus, and Poisson's ratio.

X-ray fluorescence (XRF – Bruker Hanheld Spectrometer) was used to investigate elemental composition of mudrocks. XRF was performed perpendicular and parallel to bedding plane, measurements were not found to significantly vary with bedding plane orientation, thus an average was taken. Mineral fractions are calculated using a stoichiometric method with chemical formulas of expected minerals to be found. Table 2 summarizes gravimetric content of minerals including kaolinite, illite, calcite, dolomite and quartz. Figure 3f shows shale mineral composition in a ternary diagram grouping major mineral groups.

**3.3 Scratching Procedure and Data Analyses**

An applied normal force of 31.1 N was used on all samples with the exception of soda lime glass which used 55.6 N. This normal force ensured proper scratch depth and material failure. All scratches performed in this investigation were made in the direction of the bedding plane and scratches were made at a constant speed of 0.1 mm/s. Scratches are at least 5 mm away from sample edges. Scratch length is about 5 mm. This length was sufficient to observe consistent failure of the material along the scratch path. Data acquisition recorded signals at 10 Hz sampling rate.

The procedures to calculate brittleness (detailed in Section 4.3) was translated into an algorithm and implemented in Matlab. A built-in function is used to find local minima and delimit the region of brittle effects. A minimum distance between minima is specified as a low-pass filter to ignore experimental noise. A trapezoidal rule is used to compute numerical integrals.

**4. RESULTS AND ANALYSES**

**4.1 Soda Lime Glass and Polycarbonate**

Micro-scratch tests on soda lime glass and polycarbonate were performed six times each. Results show similar trends to the ones shown in Figure 1 and are available as supplementary material (SM1 and SM2). Transverse force signatures in soda lime glass show a characteristic saw-tooth shape with large relative differences between peaks and troughs. The scratch surface is characterized by sharp wedges result of material that chipped away from the scratch center. Often, broken pieces fly away because of significant strain energy rapidly released. Transverse force signatures in polycarbonate show gradual initial increase of force as the scratch deepens towards an asymptotic limit with small relative variation of transverse force from peaks to troughs. The scratch residual looks like a continuous and smooth wedge. Failed material remains affixed to the surface and bulks out towards the edges of the scratch surface.

### 4.2 Mudrocks

Micro-scratch test on organic mudrock samples were performed four times for every sample with the exception of Vaca Muerta and Mancos which were tested five times. Figure 3 shows examples of the measured transverse force along the scratch path for each mudrock with their corresponding stereomicroscope image. Transverse force measurements showed to be quite repeatable for each mudrock sample (all measurements available as supplementary material in SM3 to SM7). In general, all transverse signatures in mudrocks follow a general trend of increasing relative distance between peak and troughs compared to the absolute force magnitude with increasing brittleness (as judged from scratch surfaces after testing). In these images, brittle failure is characterized by sharp failure surfaces while ductile failure is characterized by plastic flow around the stylus tip. In some instances, however, large variations of transverse force may develop in ductile mudrocks when the stylus intercepts a tough heterogeneity. Table 3 summarizes scratch statistics taken for transverse force data after 0.5 mm along the scratch path.

### 4.3 Micro-Scratch Brittleness

In order to estimate brittleness from the transverse force signature, we calculate the work dissipated during scratching. From an energy point of view, a brittle material accumulates strain energy through elastic deformation and releases most of it quickly upon failure (Figure 4a). On the contrary, a ductile material has a limited capacity to store strain energy and dissipates work gradually through plastic strains. Similarly, scratch testing exhibits rapid energy dissipation events which can be associated with brittle failure as evidenced from scratch surface residuals. Let us define a brittleness index $BI_W$ based on energy balance as the ratio between extra work spent in brittle events $W_B$ and total energy required for scratching $W_T$,

$$BI_W = \frac{W_B}{W_T} \qquad (1)$$

Figure 4b shows a schematic representation for an idealized transverse force-position signal. The area underneath the spikes and above the base force represent the extra work that is stored as strain energy and rapidly released upon brittle chipping, until the stylus tip catches a new edge. The process resembles stick-slip behavior in which a surface has a static friction angle higher than the dynamic friction angle. Eq. 1 may be utilized in scratch testing employing a fixed scratch depth rather than a fixed vertical load. Figure 4c shows an example of signal processing for data shown in Figure 3a. Unlike an idealized force-position signature, actual data displays variations (due to influence of local heterogeneity, noise, and mechanical vibrations) that make the selection of rapid force relaxation events dependent on the force variation amplitude or wavelength. We set a limit on the variation wavelength so that local minima are separated by a fixed distance $L_{min}$.

Table 3 summarizes the brittleness index for all scratches performed in this study. The same low-pass filtering variable $L_{min}$ is utilized for each material. Figure 5a presents the calculated mean brittleness index for all materials (corresponding error computed as one standard deviation). Brittleness indices for polycarbonate and soda lime glass are plotted as bounds for expected ductility/brittleness. As expected, polycarbonate yields the lowest brittleness index $BI_w$ < 0.04. Error bar is small due to material homogeneity and good repeatability in comparison to other specimens tested. Sodalime glass exhibits a brittleness index $BI_w$ ~ 0.19 ± 0.03. A value of $BI_w$ = 0.19 does not mean that only 19% of energy is released in brittle events. A percentage of 19% is just the extra strain energy released in brittle events since always $F_T$ > 0 (See section 5.3). Using our formulation of brittleness mudrock samples may be ranked as follows in order of increasing brittleness: Woodford, Eagle Ford, Marcellus, Mancos, Vaca Muerta (Figure 5a). This ranking is applicable to samples from specific locations and not to the entire geological formation from where samples originate.

The scratch brittleness index $BI_w$ is based on transverse force variability. Hence, a simpler estimate of britlleness may be sought through the coefficient of variation of the transverse force $F_T$ signal. Figure 5b shows the scratch brittleness index $BI_w$ plotted as a function of the coefficient of variation of $F_T$ for each scratch experiment (statistics taken after 0.5 mm of scratch length). The two quantities correlate positively, as expected, but the coefficient of variation may be affect by signal drift or material heterogeneities along the scratch path.

## 5. DISCUSSION

### 5.1 Brittleness Index v.s. Surface Residual Analysis

Inspection of the failure surface permits assessing brittleness subjectively.

<u>Soda lime glass</u>: Visual observation of soda lime glass scratch image showed clear fracture planes (Fig. 1-b). Large clear fractures appear towards the edges of the scratch while small fractures and glass comminution are characteristic towards the center. Soda lime glass brittleness index was high ($BI_W$ = 0.1949 ± 0.037) but smaller than those of the most brittle mudrocks.

<u>Polycarbonate</u>: Scratches on polycarbonate do not show sharp edges (as expected from fractures) and evidence plastic deformation and flow around the stylus tip (Fig. 1-a). Polycarbonate scratch tests resemble characteristics associated with ductile ploughing deformation mode (Briscoe et al., 1996, Bucaille et al., 2005). The absence of fracture events yields a flat transverse force signature which results in a small brittleness index ($BI_W$ = 0.0289 ± 0.0081). Small fluctuations in the force signature are the result of noise, mechanical vibration and perhaps stick slip behavior (this latter also observed in scratching polymers such as polypropylene - Jiang et al., 2015).

Mudrocks: Comparison images of scratches indicates stark contrast in failure modes as well. The Woodford scratch images are considerably smoother than any other mudrock investigated. Woodford brittleness is the lowest of all mudrocks ($BI_W$ = 0.1205 ± 0.0295) and is consistent with visual observation of scratch images. The Eagle Ford sample shows a particular stress-dependent response with ductile deformation at the stylus tip (high effective stress) and fracturing on the scratch edges (low effective stress). In comparison to Woodford, Eagle Ford has stronger evidence of chipping. The brittleness index of Eagle Ford ($BI_W$ = 0.1743 ± 0.0169) is consistent with observations of ductile deformation at the stylus tip. Marcellus shale showed the clearest visual patterns of brittle fracturing. Microfractures clearly propagate away from the main scratch path. The brittleness index of Marcellus ($BI_W$ = 0.2016 ± 0.0158) is intermediate compared to other shale samples. Mancos and Vaca Muerta scratch images indicate clear chipping but, in addition, they exhibit rough fracture planes likely caused by rock grain structure. Mancos and Vaca Muerta presented the highest brittleness index values and highest variability as well. Overall, the brittleness index $BI_W$ performed well to determine brittleness compared to surface inspection (Fig. 5).

**5.2 Brittleness Index v.s. Mineralogical Composition**

Figure 6 presents brittleness index versus weight fraction of selected major minerals. The brittleness index correlates neagatively with quartz content, is relatively neutral to total clay content and total carbonate. Dolomite content shows the highest positive correlation with the brittleness index. Although some trends appear to emerge, there is no definite correlation of brittleness index and mineral weight fractions.

Our results are in contrast to the correlation proposed by Jarvie et al. (2007),

$$BI_Q = \frac{Q}{Q+C+Cl} \tag{2}$$

regarding the influence of quartz on brittleness, where $Q$ is the amount of quartz, $C$ is the amount of total carbonate, and $Cl$ is the total amount of clay (Fig. 6). Other studies also showed poor correlations of failure-assessed brittleness with $BI_Q$ (Yang et al., 2013). Such brittleness index $BI_Q$ was developed for Barnett shale and may not be applicable to other organic mudrocks. Our results agree with others in that the presence of dolomite increases brittleness (Wang and Gale, 2009). In summary, mineralogy without any other information seems insufficient to determine brittleness for various shales with distinct origin.

### 5.3 Brittleness Index v.s. Dynamic Elastic Coefficients

Correlations based on elastic coefficients attribute high brittleness to rocks with a large Young's modulus and small Poisson's ratio. The following equation captures such conjecture:

$$BI_E = \frac{1}{2}\left(\frac{E[10^6 \text{ psi}]-1}{8-1} + \frac{\nu-0.4}{0.15-0.4}\right) \quad (3)$$

where $E$ is the rock Young's modulus and $\nu$ is the Poisson ratio (Rickman et al., 2008). Such equation considers $1\cdot10^6$ psi and $8\cdot10^6$ psi as limits for Young's modulus and 0.15 and 0.4 as limits for Poisson's ratio. Figure 7a shows a fair positive correlation of scratch $BI_W$ with respect to $BI_E$. There is no consistent and unique physical link between small strain elastic properties (E, ν) and failure properties such as brittleness. Yet, a high Young's modulus is an indicator of large elastic strain energy accumulation during loading and a low Poisson ratio often indicates a well cemented rock matrix opposite to mudrocks that may exhibit little volumetric strain change when loaded deviatorically (high apparent ν).

The $BI_E$ index places in the same category Marcellus and Woodford which exhibit marked differences in deformation modes. Porosity seems a missing parameter to constrain brittleness predictions from mineralogical or elastic properties. Fig. 7b shows the scratch brittleness index $BI_W$ as a function of material mass density, as a proxy for porosity. Results show a well-defined trend of increasing brittleness with increasing mass density. Woodford scratch brittleness appears separated from other tested shales likely as a result of higher porosity.

### 5.4 Other Scratch Brittleness Calculations

Brittleness may also be measured in scratch testing using the 3-response model of deformation (Krupicka et al., 2003). However, the 3-response model is essentially only used to investigate wear resistance of polymer coatings (Barletta et al., 2007; Barletta et al., 2013; Krupicka et al., 2003). Polycarbonate has been found to hardly fail in a brittle mode with a brittleness of 0.25% to 3% (Barletta et al., 2013). This model is time intensive and difficult to apply to rough scratches with substantial material removal, as in the case of mudrocks. Though our model agrees that polycarbonate is a low brittleness material, mudrocks have not been investigated using the 3-response model and the consistency of this model with ours is beyond the scope of this investigation.

**5.5 Experimental Conditions**

Variation of experimental conditions can affect scratch patterns and brittleness calculation. The following list summarizes main factors to take into account for obtaining consistent brittleness estimations.

<u>Scratch speed</u>. Strain-rate dependency of rock deformation characteristics is well-known. High strain-rates promote brittle failure. Scratch speed was not varied in our investigation. However scratch speed is known to effect scratching response for some materials. In soda lime glass, low scratch speeds cause micro-cracks and fractures to evolve and widen a scratch, while high speeds may induce shear failure (Bandyopadhyay, 2012). Hence, we recommend using low scratch speed to resemble quasi-static deformation and allow fractures –if any– to propagate.

<u>Vertical load and scratch depth</u>: Applied load was not varied in our investigation. The effects of applied load have in soda lime glass show that the appearance and degree of micro-chip formation vary with applied load, and transition zone failure and brittleness may be affected (Bandyopadhyay, 2012). Our findings are based on applied loads which were found to yield suitable and repeatable failure characteristics. From all mudrocks, Eagle Ford showed the highest failure surface variability along the scratch depth. Deeper scratch depth would likely could change such behavior.

<u>Constant vertical load and constant scratch depth</u>: We validated our scratch brittleness formulation for scratch experiments with constant vertical load. Yet, the same type of analysis would be suitable to constant scratch depth experiments, since brittle failure would yield large variations of transverse force (Richard et al., 2012). The validation of such procedure is outside the scope of this manuscript.

<u>Scratch frame stiffness</u>: The stiffness of the scratch device is critical to capturing brittle events. A large magnitude of transverse force $F_T$ may be accumulated in the scratch frame and lead to erroneous estimations of brittleness. We believe this is the reason for which scratch glass brittleness seems to be low in our study. Recent numerical simulations show that in an infinitely rigid frame, transverse force experiences large variations and may approach zero after a brittle failure event (Sun et al., 2017). Imperfections in stylus bearings may also cause tip slippage and lateral movement and result in transverse force variations not related to brittleness.

**5.6 Influence of Rock Heterogeneity at Various Length Scales on Brittleness**

The micro scratch is useful for investigating brittleness of the rock matrix and spatial variations at scales smaller than ~10 mm (e.g., identification of finely laminated shales - Akono and Kabir, 2015). Yet, micro scratch brittleness may not directly upscale to assess formation brittleness in views of hydraulic fracturing. Formation heterogeneity at reservoir scale may dominate fracture propagation and greatly contribute to increasing surface area during hydraulic fracturing at large scales (Bing et al., 2014, Warpinski, 2008). Such heterogeneities include foliation, lamination/bedding, and natural fractures (Gale 20XX; Warpinsky et al., 2008; Zhou et al., 2008; Guo et al., 2017).

Supplementary figure SM8 shows an X-ray microtomography image of the Eagle Ford field core from which we cut a sample for micro scratch testing. Based on inspection of SM8, variations in constituency and cracks are present in the field core sample. However the bulk is dominated by homogenous rock matrix. Other rocks samples may present much stronger heterogeneity. Triaxial experiments combined with X-ray microtomography show that even relatively soft rocks ($E$ < 10 GPa) can develop large surface area upon loading when natural fractures and marked lamination are present (Espinoza et al., 2015, 2016).

We propose to use micro scratch brittleness to enrich results from well logs and field-core scratching, and to be used as alternative when other methods are inaccessible (e.g., when reservoir cores are not available but drill cuttings are).

## 6. CONCLUSIONS

Mudrock matrix brittleness is an important parameter to help assess effectiveness of fracturing to create new surface area. We developed a method to quantify brittleness utilizing the micro scratch test. The method is based on actual rock failure phenomena rather than mineralogical correlations or small strain elastic parameters.

Brittleness index results agree well with observation of the surface residuals after scratching and expectations from elastic parameters. We did not find a definite correlation with mineralogy. Yet, results from various types of organic mudrocks indicate a decrease of brittleness with increasing presence of quartz and an increase of brittleness with increasing dolomite content. Porosity seems to be another variable that affects brittleness.

This study focuses on rock matrix brittleness. Other factors that may influence creation of new surface area upon fracturing include rock heterogeneity and presence of natural discontinuities and fractures. Rock matrix brittleness, as determined with the micro scratch tests, may supplement current laboratory tests on rock plugs and complement well-log measurements and field core scratch testing.

**Acknowledgments**

This research was supported by the Petroleum and Geosystem Engineering Summer Undergraduate Research Internship at the University of Texas at Austin. M. D. Aman was supported by the Center for Frontiers of Subsurface Energy (CFSES) under Contract No. DE-SC0001114 with the U.S. Department of Energy, Office of Science, Office of Basic Energy Sciences. We thank A. G. Ilgen, A. Sosa-Massaro, S. Cuervo and M. M. Sharma for providing some of the tested samples and access to XRF equipment.

# Tables and Figures

**Table 1. Samples tested with micro scratch test.**

| Sample | Type of Core | Geologic Context | Mass density [kg/m$^3$] | Dynamic Young's modulus [GPa] | Dynamic Poisson's ratio [-] | Number of scratches |
|---|---|---|---|---|---|---|
| Soda lime Glass | NA | NA | 2650 | 63.4 | 0.17 | 6 |
| Polycarbonate | NA | NA | 1226 | 2.3 | 0.40 | 6 |
| Eagle Ford | Field Core | Cretaceous | 2493 | 57.0 | 0.13 | 4 |
| Mancos | Outcrop | Cretaceous | 2558 | 24.1 | 0.08 | 5 |
| Marcellus | Outcrop | Devonian | 2476 | 20.7 | 0.22 | 4 |
| Vaca Muerta | Field Core | Jurassic – Cretaceous | 2540 | 44.2 | 0.16 | 5 |
| Woodford | Outcrop | Devonian | 2087 | 18.3 | 0.19 | 4 |

**Table 2. Mineral weight content of samples tested obtained from XRF analysis**

| Sample | Quartz (%) | Kaolinite (%) | Illite (%) | Total Clay (%) | Calcite (%) | Dolomite (%) | Total Carbonate (%) | Others (%) |
|---|---|---|---|---|---|---|---|---|
| Eagle Ford | 5.8 | 3.2 | 7.5 | 10.7 | 73.9 | 1.7 | 75.6 | 7.9 |
| Mancos | 35.9 | 10.9 | 25.4 | 36.3 | 8.9 | 6.2 | 15.2 | 12.6 |
| Marcellus | 39.5 | 11.1 | 25.7 | 36.8 | 4.2 | 2.7 | 6.9 | 16.8 |
| Vaca Muerta | 7.5 | 9.3 | 21.7 | 31.0 | 32.4 | 24.2 | 56.7 | 4.8 |
| Woodford | 52.9 | 9.3 | 21.7 | 31.0 | 1.3 | 2.4 | 3.7 | 12.4 |

**Table 3. Summary of Scratch Statistics and Brittleness Index.**

| Sample | Average ($F_T$) [N] | Standard Deviation ($F_T$) [N] | $L_{min}$ [mm] | Brittleness Index $BI_W$ [J/J] |
|---|---|---|---|---|
| Soda lime glass | 5.5982 | 0.7591 | 0.40 | 0.1630 |
| | 5.6805 | 0.8174 | | 0.2057 |
| | 5.5751 | 0.7790 | | 0.1896 |
| | 6.4619 | 0.9104 | | 0.1817 |
| | 5.6583 | 0.7554 | | 0.1789 |
| | 5.2561 | 0.8099 | | 0.2506 |
| Polycarbonate | 15.0447 | 0.4326 | 0.20 | 0.0307 |
| | 14.7239 | 1.0078 | | 0.0252 |
| | 13.9820 | 0.7710 | | 0.0298 |
| | 13.9593 | 0.5729 | | 0.0256 |
| | 13.1498 | 0.4601 | | 0.0192 |
| | 13.0346 | 0.7296 | | 0.0431 |
| Eagle Ford | 12.1897 | 1.1865 | 0.35 | 0.1554 |
| | 11.6092 | 1.0472 | | 0.1788 |
| | 11.9327 | 1.2748 | | 0.1953 |
| | 11.4971 | 1.0069 | | 0.1678 |
| Mancos | 16.1483 | 1.9754 | 0.45 | 0.1523 |
| | 16.0564 | 2.2996 | | 0.1628 |
| | 10.7549 | 2.0355 | | 0.2259 |
| | 14.7498 | 3.0489 | | 0.2935 |
| | 12.8906 | 1.9089 | | 0.1997 |
| Marcellus | 14.2562 | 2.2465 | 0.37 | 0.2044 |
| | 14.1681 | 1.7759 | | 0.1820 |
| | 14.5829 | 3.0233 | | 0.1998 |
| | 13.9445 | 2.2831 | | 0.2203 |
| Vaca Muerta | 14.0341 | 2.2363 | 0.40 | 0.2368 |
| | 14.1696 | 2.4151 | | 0.1876 |
| | 13.3776 | 2.0393 | | 0.2582 |
| | 13.6138 | 2.4275 | | 0.2761 |
| | 12.1259 | 2.3458 | | 0.3107 |
| Woodford | 12.9075 | 1.5729 | 0.27 | 0.1493 |
| | 12.8449 | 1.3192 | | 0.0958 |
| | 12.4251 | 1.3838 | | 0.1426 |
| | 12.4478 | 0.9511 | | 0.0941 |

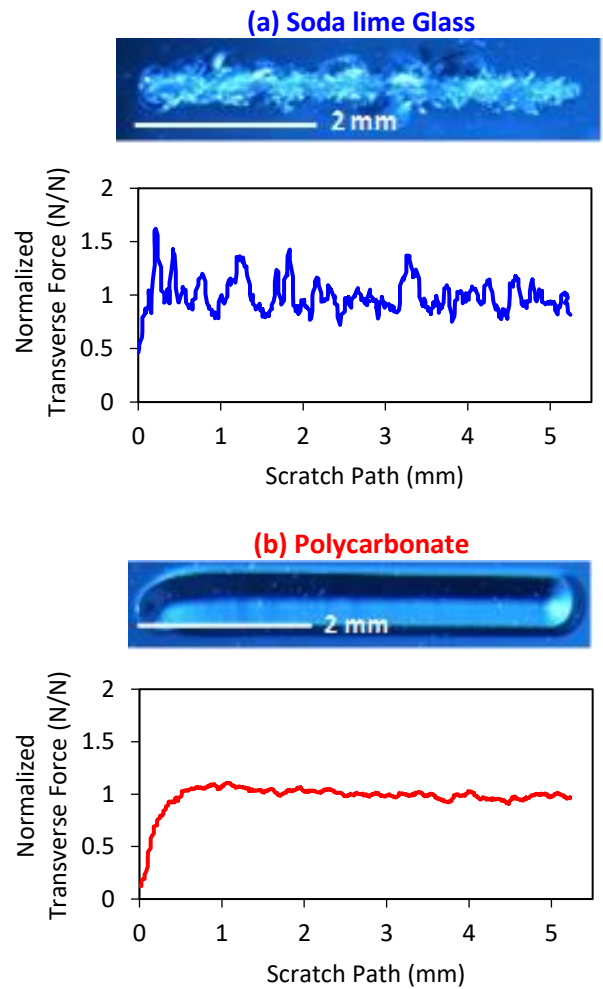

Figure 1. Scratches on soda lime glass (a) and polycarbonate (b). Transverse force signatures (normalized by plateau average) and scratch residuals highlight the difference between scratching a brittle (soda lime glass) and ductile (polycarbonate) material.

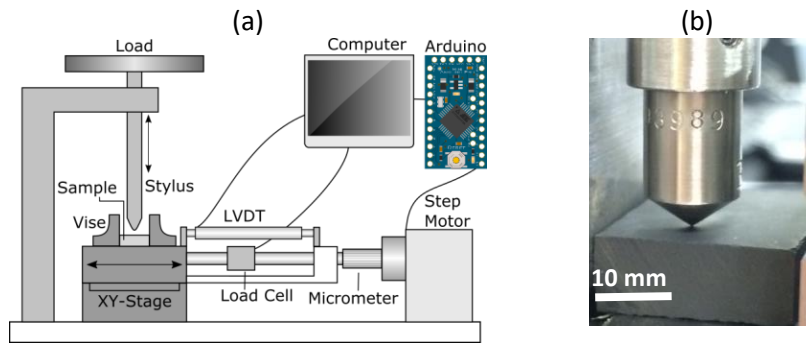

**Figure 2. (a) Micro-scratch test assembly. (b) Conical-spherical tip in contact with a mudrock sample.**

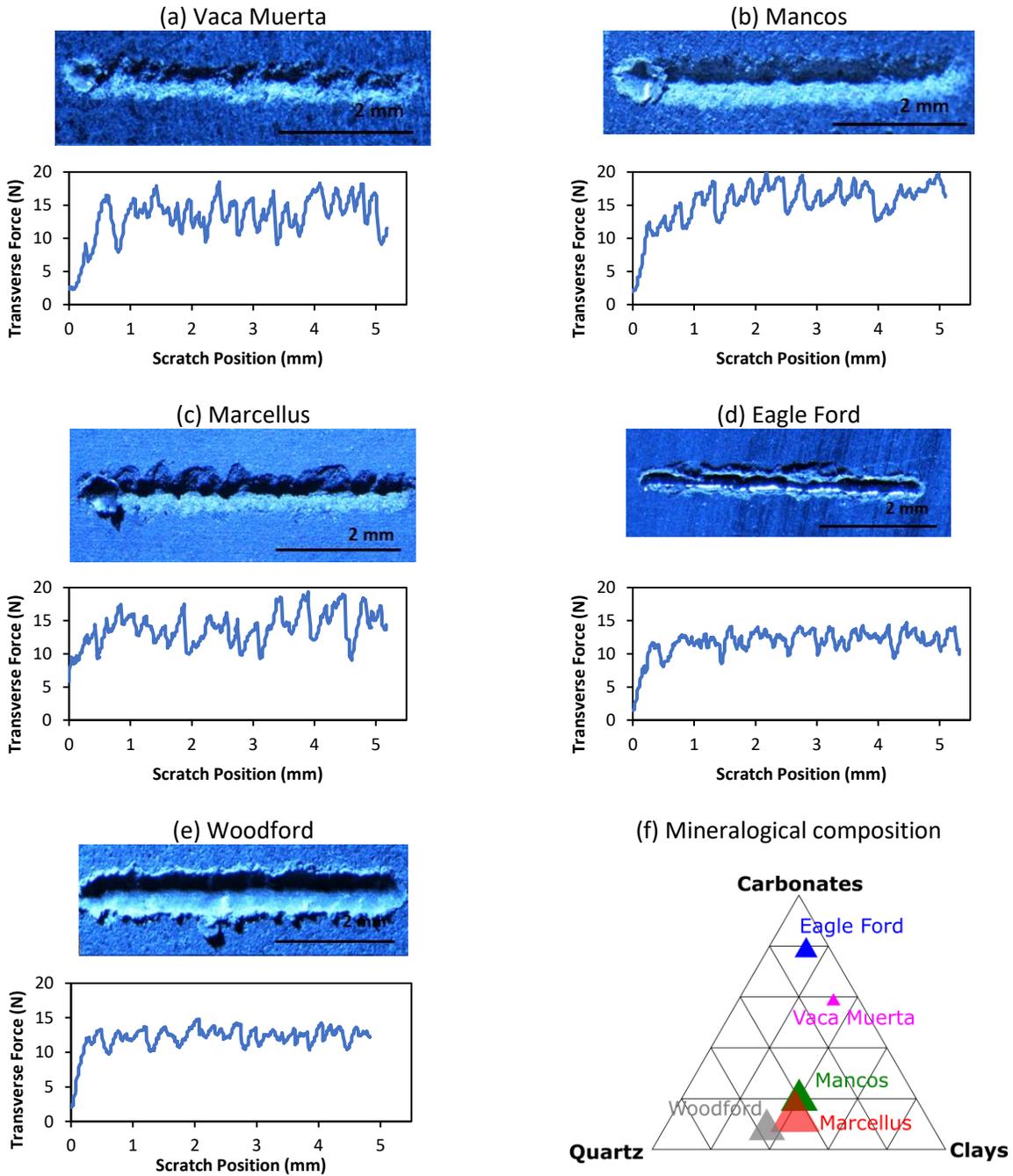

Figure 3. Examples of transverse force v.s. stylus position measured during scratching. Rock samples are organized from most brittle (a) to least brittle (e) according to Figure 5. (f) Ternary diagram with shale mineralogical composition.

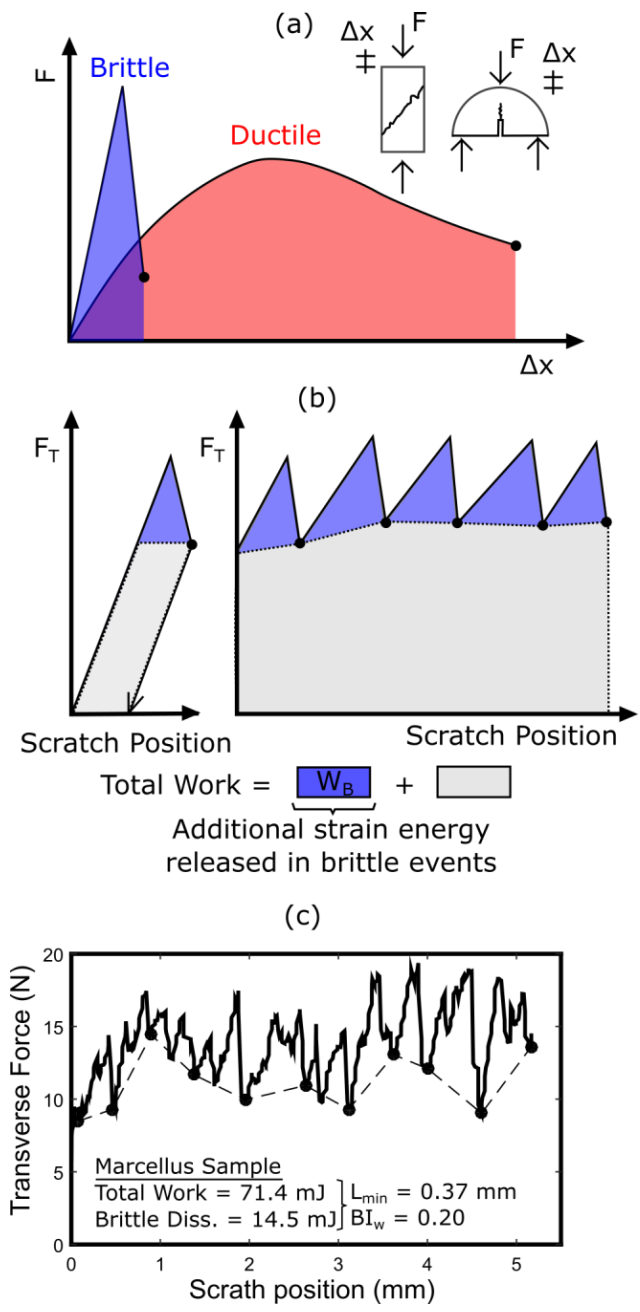

**Figure 4.** (a) Strain-rate based comparison between brittle and ductile deformation and failure. (b) Extension of strain-rate based brittleness to the scratch test. (c) Example of brittleness calculation for data from Fig. 3a.

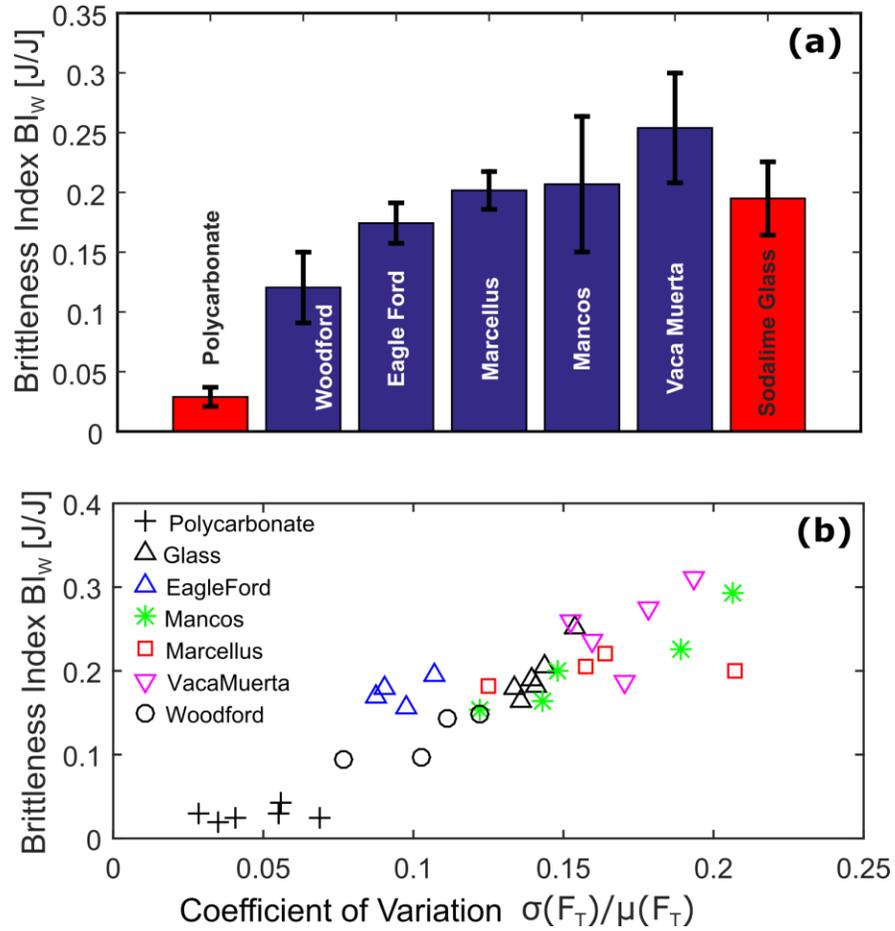

Figure 5. (a) Summary of scratch brittleness index for various materials tested. Error bars correspond to one standard deviation. (b) Scratch brittleness index as a function of scratch transverse force variability (σ: standard deviation, μ: mean – statistics after 0.5 mm of scratching).

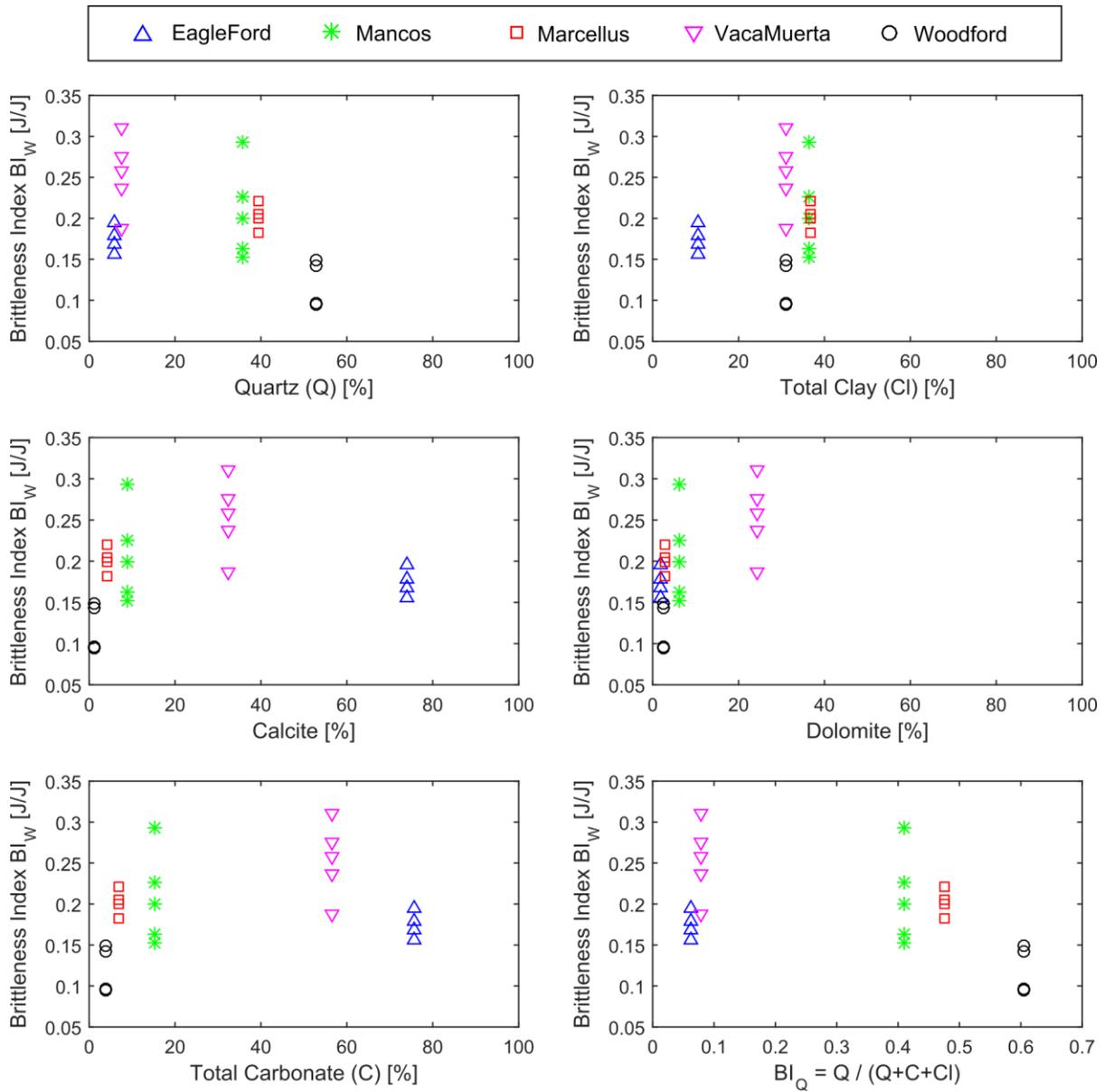

Figure 6. Brittleness index as a function of weight fraction of selected major mineral groups, and as a function of brittleness index based on mineralogy (Eq. 2).

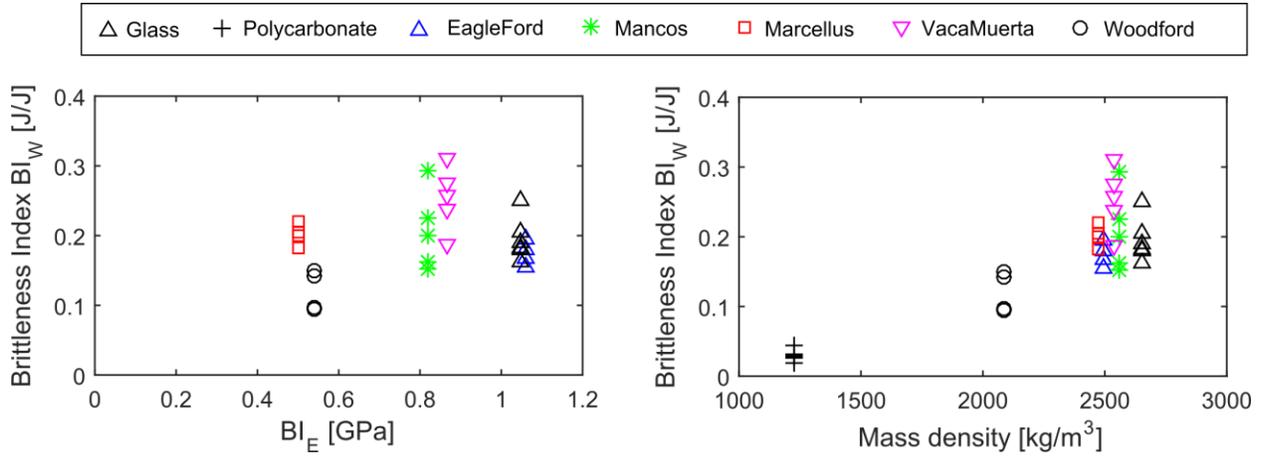

Figure 7. Brittleness index plotted as a function of brittleness index BI$_E$ based on Young's modulus over Poisson ratio – Eq. 3 (a), and as a function of material mass density (b).

**Supplemental Material**

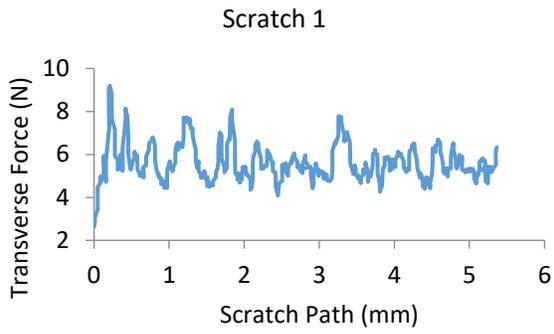
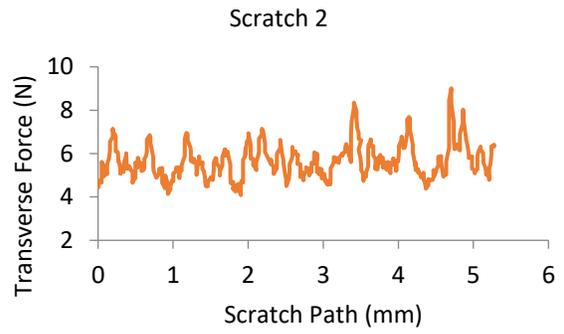
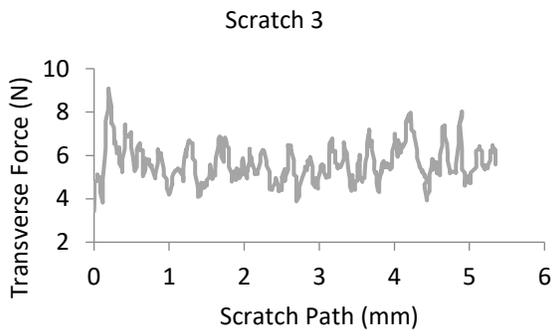
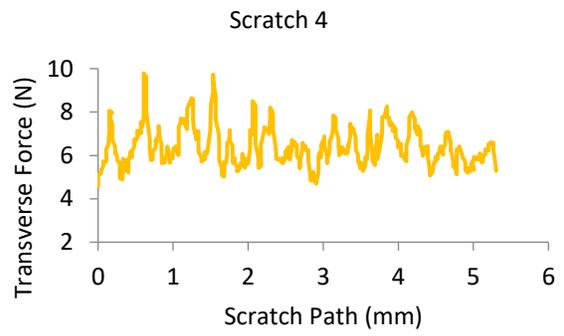
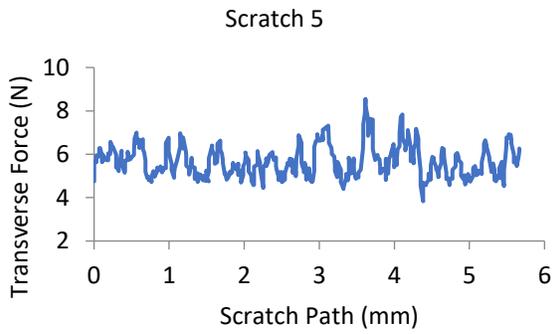
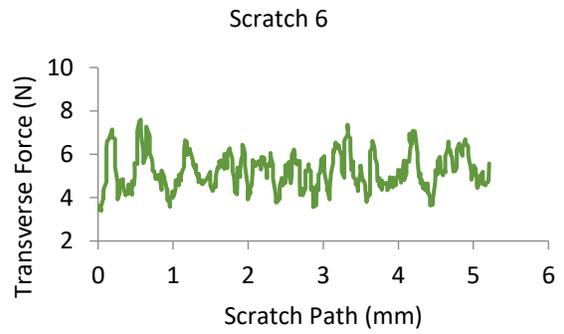

**SM1. Transverse Force-Scratch Path signatures for Soda lime Glass Scratches**

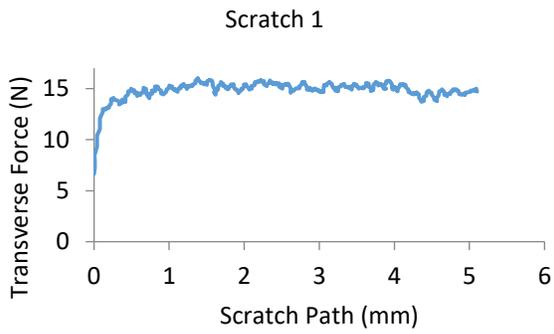
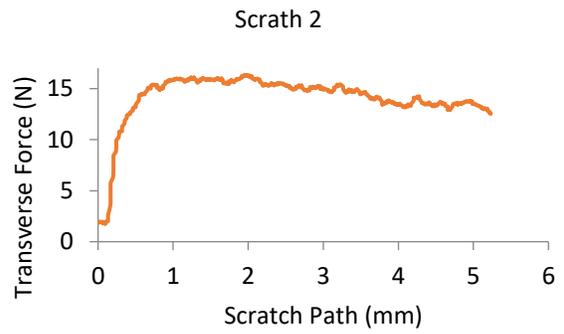
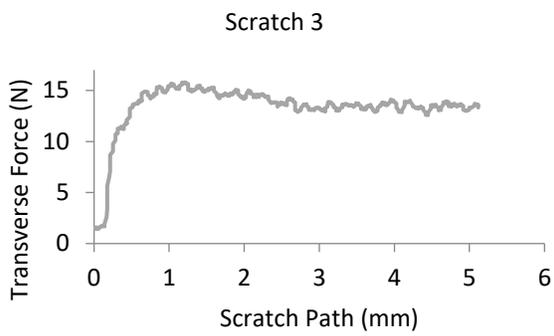
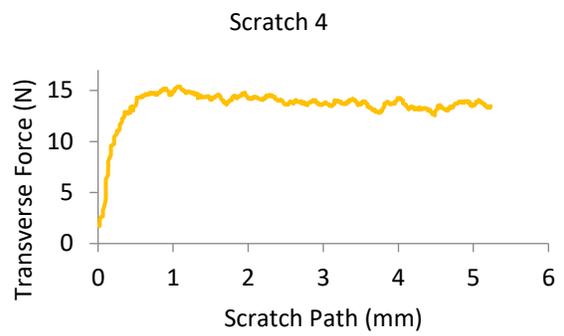
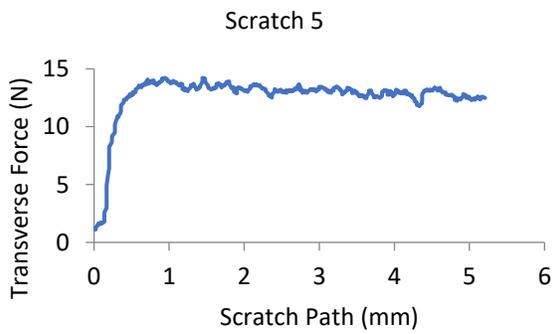
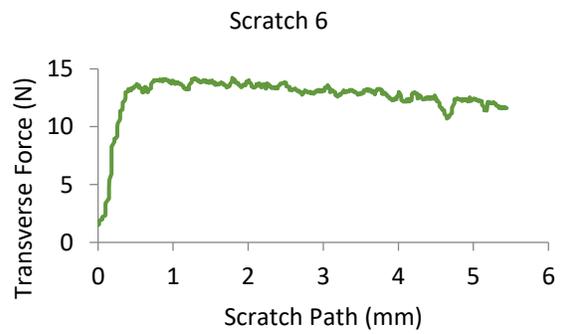

**SM2. Transverse Force-Scratch Path signatures for Polycarbonate Scratches**

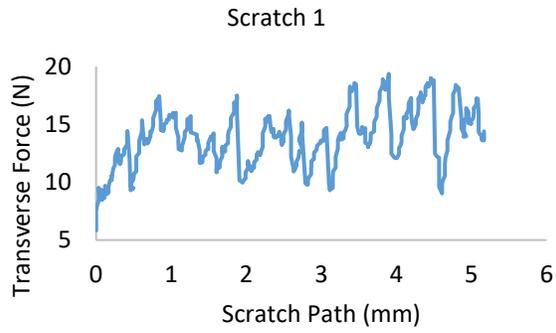
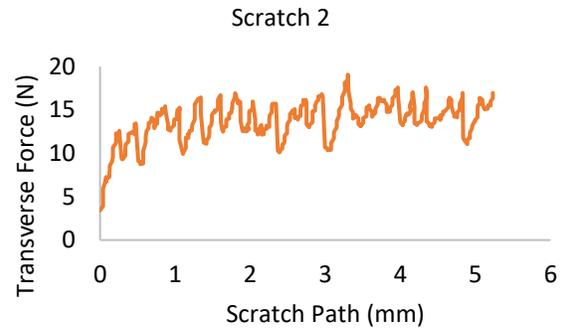
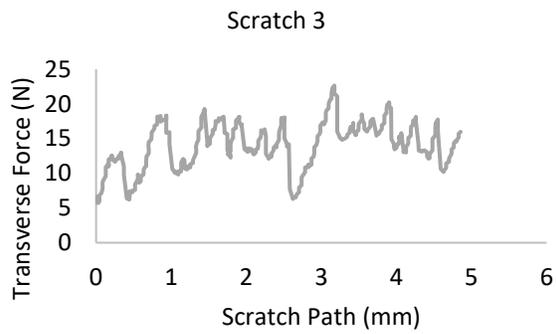
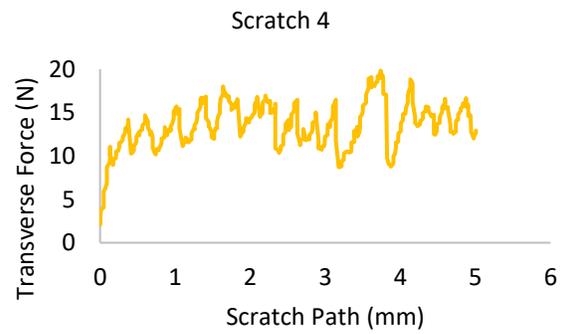

**SM3. Transverse Force-Scratch Path signatures for Marcellus scratches**

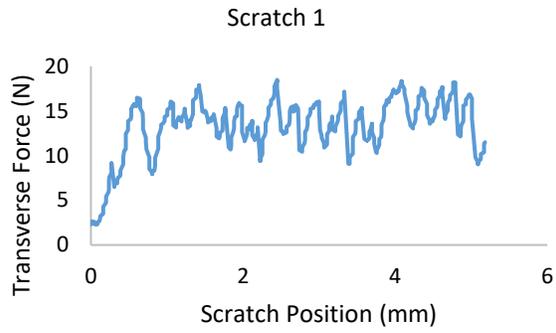
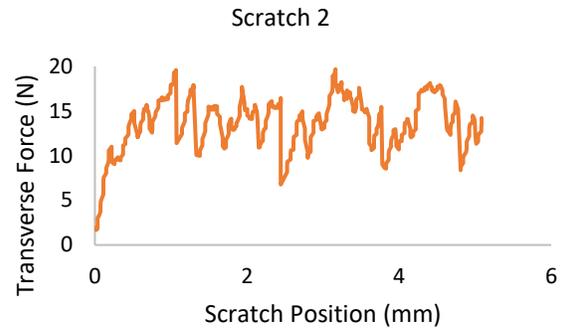
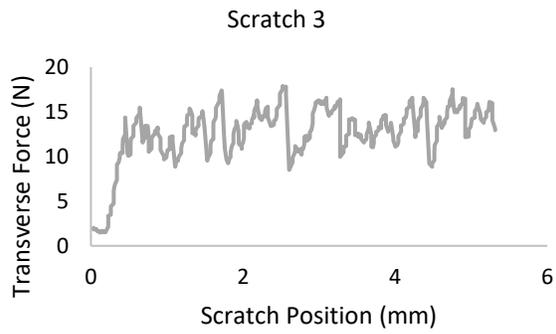
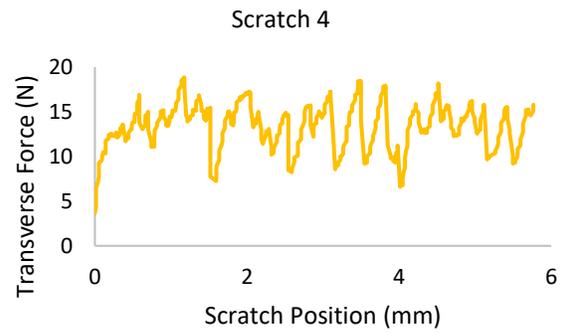
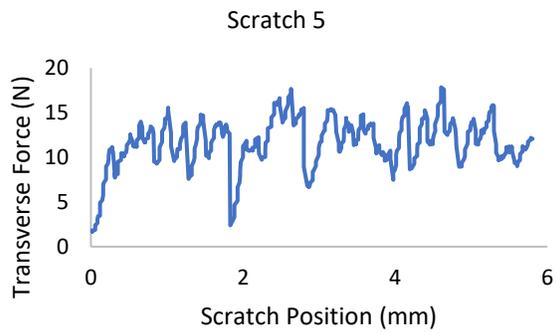

**SM4. Transverse Force-Scratch Path signatures for Mancos scratches**

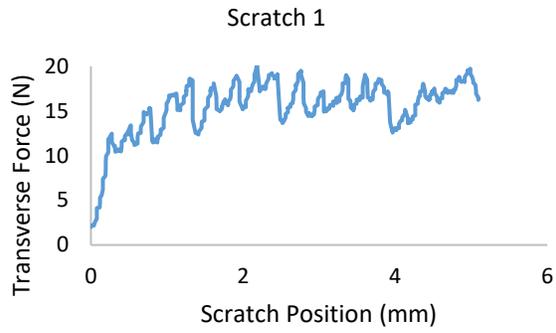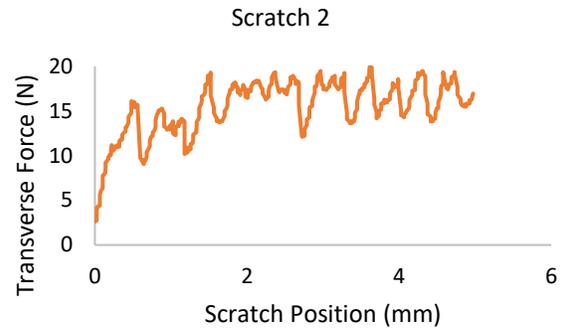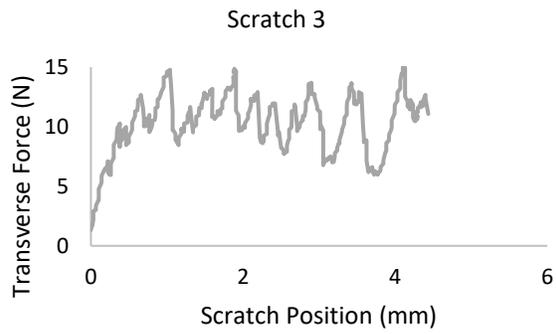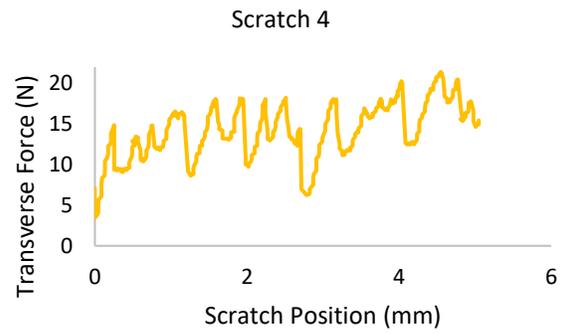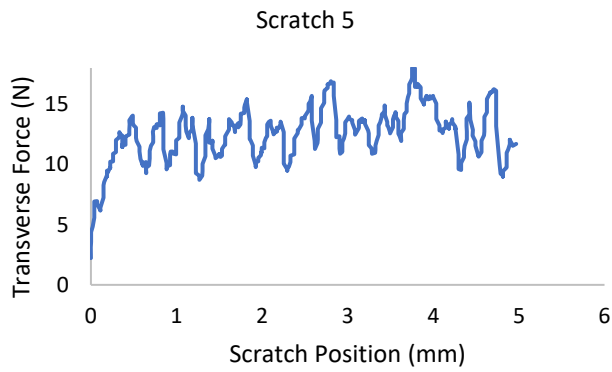

**SM5. Transverse Force-Scratch Path signatures for Vaca Muerta scratches**

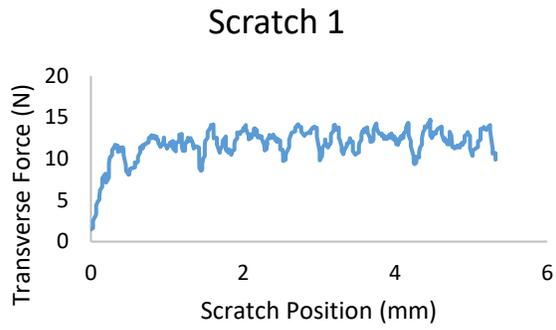
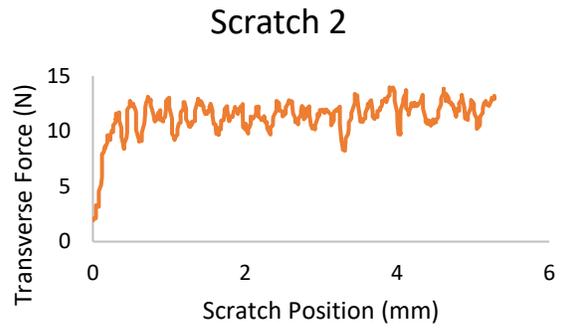
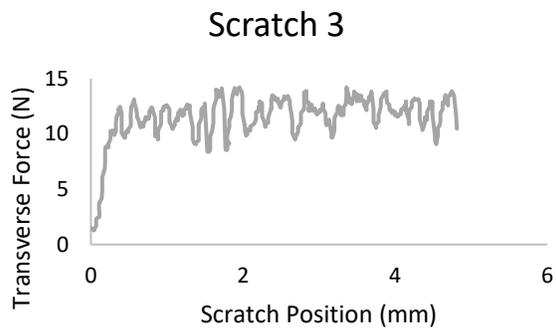
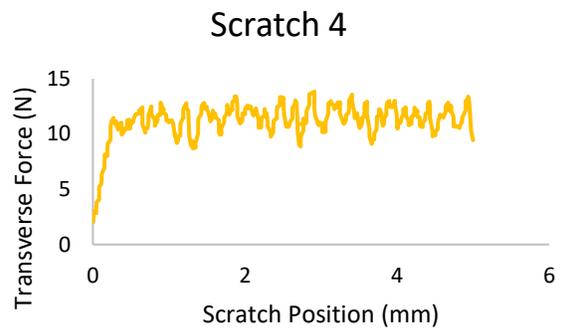

**SM6. Transverse Force-Scratch Path signatures Eagle Ford scratches**

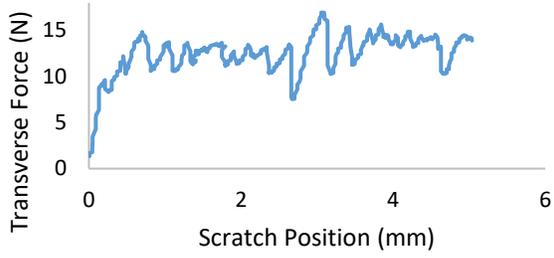
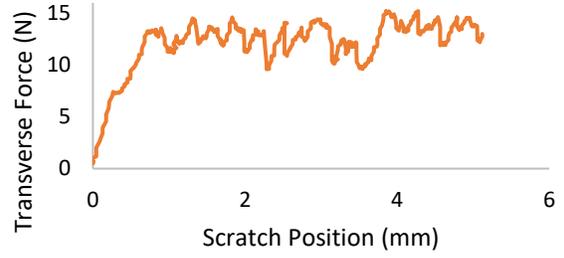
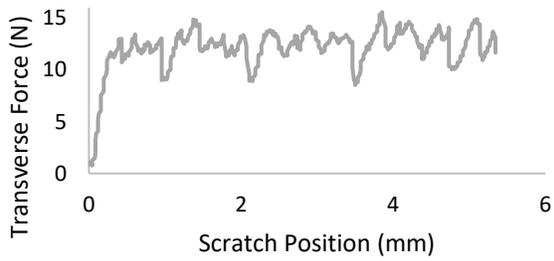
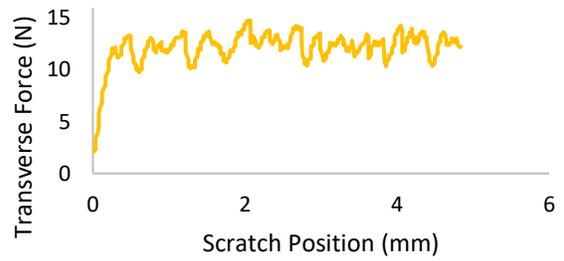

**SM7. Transverse Force-Scratch Path signatures Woodford scratches**

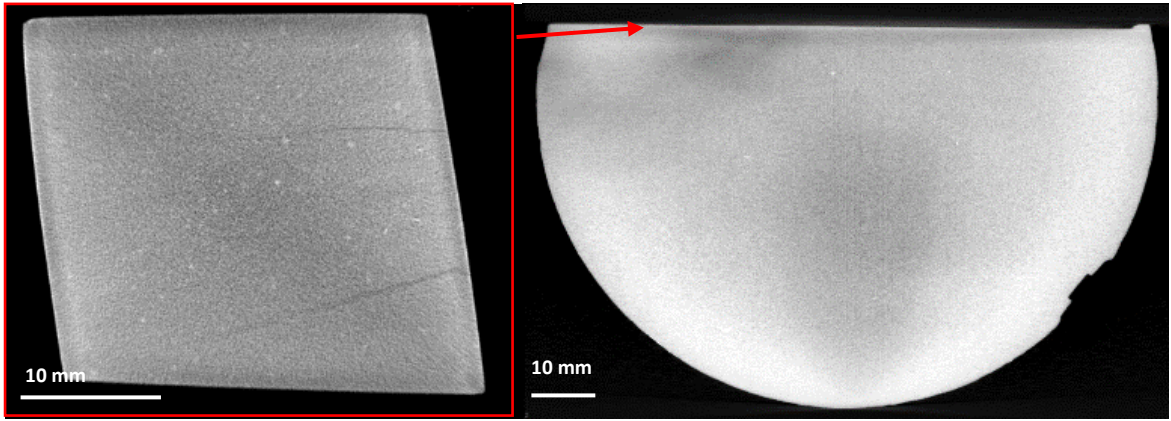

**SM8. Micro CT Images of Eagle Ford field core**